\documentclass[doublecol]{epl2}

\title{Noise-induced interspike interval correlations and
spike train regularization in spike-triggered adapting neurons}

\shorttitle{Interspike interval correlations of adapting neurons
in subthreshold regime}

\author{Eugenio Urdapilleta\inst{1}}
\shortauthor{E. Urdapilleta}

\institute{
  \inst{1} Divisi\'on de F\'isica Estad\'istica e Interdisciplinaria,
  Centro At\'omico Bariloche, Av. E. Bustillo Km 9.500, S. C. de
  Bariloche (8400), R\'io Negro, Argentina
}


\abstract{Spike generation in neurons produces a temporal point
process, whose statistics is governed by intrinsic phenomena and
the external incoming inputs to be coded. In particular,
spike-evoked adaptation currents support a slow temporal process
that conditions spiking probability at the present time according
to past activity. In this work, we study the statistics of
interspike interval correlations arising in such non-renewal spike
trains, for a neuron model that reproduces different spike modes
in a small adaptation scenario. We found that correlations are
stronger as the neuron fires at a particular firing rate, which is
defined by the adaptation process. When set in a subthreshold
regime, the neuron may sustain this particular firing rate, and
thus induce correlations, by noise. Given that, in this regime,
interspike intervals are negatively correlated at any lag, this
effect surprisingly implies a reduction in the variability of the
spike count statistics at a finite noise intensity.}

\begin{document}
\maketitle

\section{Introduction}
\indent In many instances, point processes describing the firing
statistics of different neurons go beyond the simple Poissonian
declaration of temporal events or the more general renewal
processes, which are often considered good descriptions of
stationary spike trains \cite{Gerstner_etal_2014,
Perkel_etal_1967, ShadlenNewsome1998}. Non-renewal firing
properties have been observed in different species and neural
areas \cite{LowenTeich1992, RatnamNelson2000, NeimanRussell2001,
Bahar_etal_2001, Nawrot_etal_2007, Farkhooi_etal_2009,
Peterson_etal_2014}. The lack of independence between subsequent
interspike intervals (ISIs), which defines the non-renewal
character of a point process, may arise from different endogenous
and exogenous mechanisms \cite{AkerbergChacron2011}. In
particular, spike-evoked adaptation currents are one of the most
prominent processes shaping the statistical structure of
non-renewal spike trains \cite{AkerbergChacron2011, LiuWang2001,
Chacron_etal_2001, PrescottSejnowski2008, Benda_etal_2010,
Schwalger_etal_2010, Nesse_etal_2010, Farkhooi_etal_2011,
Urdapilleta2011, SchwalgerLindner2013, Shiau_etal_2015}.

\indent Given that spikes constitute the main substrate for
neuronal communication \cite{Gerstner_etal_2014}, these correlated
events not only highlight the presence of certain
history-dependent processes, but also imply profound effects on
neural coding. For example, rate coding of static inputs is
strongly affected by correlations between subsequent ISIs
\cite{CoxLewis1966, VanVreeswijk2010}, as well as the information
transfer of slow signals \cite{Chacron_etal_2004,
Lindner_etal_2005}. In particular, adaptation currents generally
induce negative correlations, resulting in a long-term reduction
of the variability of the spike count statistics
\cite{AkerbergChacron2011, LiuWang2001, Chacron_etal_2001,
PrescottSejnowski2008, Benda_etal_2010, Schwalger_etal_2010,
Nesse_etal_2010, Farkhooi_etal_2011, Urdapilleta2011,
SchwalgerLindner2013, Shiau_etal_2015}, but richer patterns of
correlations are also possible \cite{SchwalgerLindner2013,
Shiau_etal_2015}.

\indent Neurons respond to incoming stimuli in different ways.
Type-I neurons or ``integrators'' are an important class of
excitable neural cells, in which input signals are integrated up
to a threshold, without any strong modulation by the spectral
characteristics of fluctuations \cite{Prescott_etal_2008,
MatoSamengo2008}. For these neurons, it is important to
differentiate two firing regimes: sub-threshold (or
fluctuation/noise-driven) and supra-threshold (or
input/mean-driven) modes \cite{Gerstner_etal_2014,
Rauch_etal_2003, LaCamera_etal_2008}. Whereas in the first regime,
neuronal dynamics has only a stable quiescent state and spiking
responses can be reached only assisted by noise, in the second
one, a repetitive firing is obtained even in a deterministic
scenario and noise simply makes that trajectories fluctuate around
a deterministic cycle. Different analytical studies have focused
on the role of adaptation currents in generating interspike
interval correlations in spike trains of neuron models set to the
supra-threshold regime \cite{Urdapilleta2011,
SchwalgerLindner2013, Shiau_etal_2015}. However, many cortical
areas exhibit a fluctuation-sensitive or balanced regime
\cite{ShadlenNewsome1998, VanVreeswijkSompolinsky1998,
VogelsAbbott2009}, typically from a sub-threshold dynamics, and
therefore, it would be important to assess the contribution of
adaptation currents on the non-renewal characteristics of spike
trains in this condition.

\indent In this work, we address this analysis for the minimal
dynamical model supporting this phenomenon: the \textit{leaky}
integrate-and-fire (LIF) model. Two main approaches have been used
to study correlations in integrate-and-fire (IF) models (although
others can also be adapted): a method in which correlations are
studied through the analysis of perturbations on a limit cycle
\cite{Schwalger_etal_2010, SchwalgerLindner2013}, and a method
derived from the formulation of an appropriate hidden Markov model
(HMM) \cite{Chacron_etal_2003, VanVreeswijk2010, Urdapilleta2011}.
The first method is extremely useful to study cases where
adaptation produces realistic conditions, but it is restricted
only to neuron models set in a repetitive firing regime, thus
preventing its application to the analysis of the sub-threshold
regime. By construction, the second method can be applied to any
situation, but useful results were obtained analytically only for
a perturbative regime of small adaptation currents
\cite{Urdapilleta2011}. Interestingly, both approaches result in
correlations with essentially the same mathematical structure.
Based on the second approach, here we study how interspike
interval correlations behave in response to different features of
the incoming stimuli, with particular emphasis in the noise-driven
regime. Since previous studies have shown that a non-trivial
structure of correlations arises as the firing rate of the spiking
neuron is changed (by manipulating the deterministic drift, or an
equivalent parameter) \cite{LiuWang2001, PrescottSejnowski2008,
Benda_etal_2010, Schwalger_etal_2010, Urdapilleta2011,
SchwalgerLindner2013, Shiau_etal_2015}, we hypothesize that a
similar situation can be reached, in the sub-threshold regime,
when noise varies, by setting the firing rate at selected values.
Confirming this hypothesis, we found that, in any sub-threshold
regime defined by a fixed drift, negative correlations are maximal
at a finite noise, implying a surprising regularizing effect of
noise on long-term spike count variability. Further studies on the
consequences of this effect are under analysis.

\section{Simplified neuron models with spike-evoked adaptation}

Single-cell neuronal models describe the electrical properties of
voltage-sensitive membranes, including their response to incoming
signals. Often, integration of signals produces stereotyped spikes
when the transmembrane potential reaches certain value or
threshold. Initiated by the foundational description of the
excitability of the squid axon by Hodgkin and Huxley in 1952,
conductance-based models characterizing the behavior of ionic
channels and their interaction with the membrane potential account
not only for this highly nonlinear process of spike generation
\cite{Gerstner_etal_2014}, but also for many other subthreshold
phenomena including oscillations \cite{HutcheonYarom2000} or
adaptation with subthreshold activation \cite{BrownAdams1980,
PrescottSejnowski2008}, among others. Unlike high-dimensional
detailed models, IF neuronal models are approximate descriptions
that relieve the need of a precise spike generation mechanism and
simply produce spikes by declaration. However, in order to keep as
much information as possible between spikes, the description of
the electrical evolution of the potential during the subthreshold
period should include all relevant phenomena. For type-I neurons,
the LIF model can be considered as the minimal model preserving
the characteristics of neuronal processing during rest (leaky
current and asymptotic relaxation).

\begin{figure}[!t]
\centerline{\includegraphics[width=0.95\linewidth]{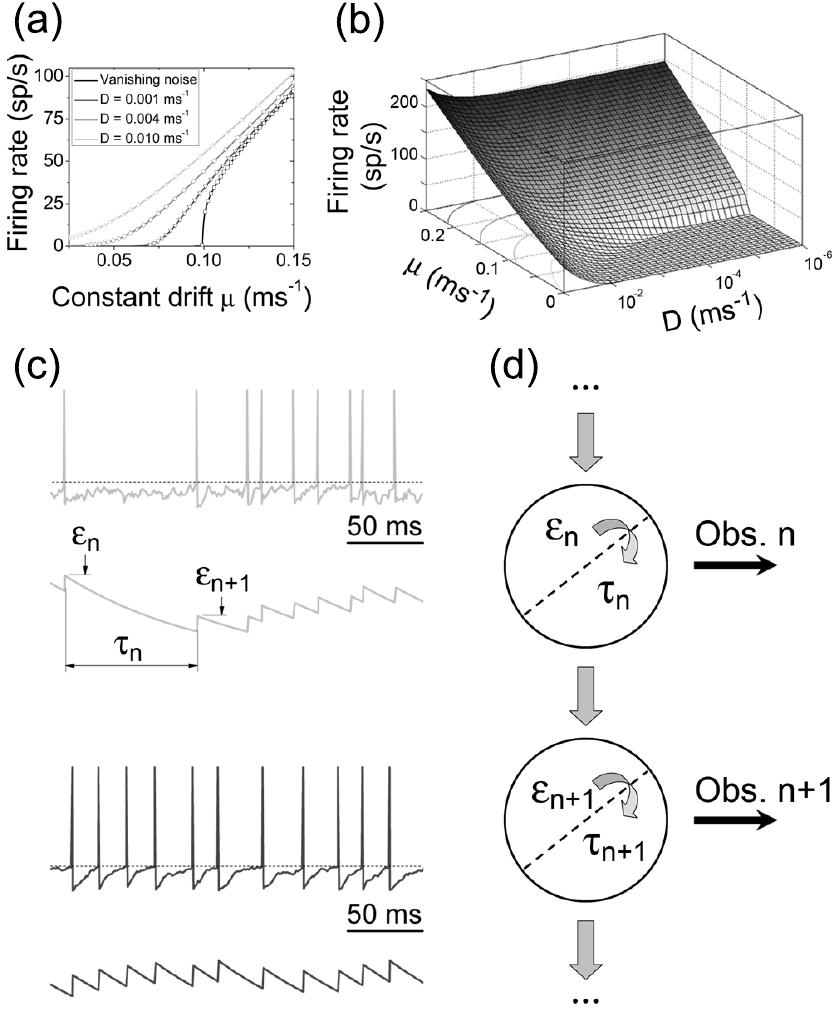}}
\caption{The LIF neuron model and the hidden Markov correlated
sequence of ISIs. (a) Firing rate response of a LIF neuron as a
function of the constant drift driving the dynamics, $\mu$, for
different noise intensities, $D$. (b) Firing rate response as a
function of $\mu$ and $D$. Parameters: $V_{\rm r} = 0$, $V_{\rm
thr} = 1$, $\tau_{\rm m} = 10~{\rm ms}$, $\alpha = 0$ (no
adaptation). (c) Simulated voltage trace and adaptation current in
the sub-threshold (upper figures, light-gray lines) and the
supra-threshold (lower figures, dark-gray lines) regimes.
Variables defining the HMM used for the analysis of correlations
between ISIs are indicated. (d) Hidden Markov model: the
statistics of the ($n+1$)-th observation (ISI), $\tau_{{\rm
n}+1}$, is conditioned to the ($n+1$)-th initial adaptation level,
$\varepsilon_{{\rm n}+1}$, which in turn only depends on both
variables at the immediately previous stage.} \label{fig.1}
\end{figure}

Neuronal adaptation includes a variety of interrelated processes,
from spike-triggered and subthreshold-activated adaptation
currents \cite{PrescottSejnowski2008} to multi-time-scale
adaptation \cite{Nesse_etal_2008, Pozzorini_etal_2013}. The
addition of a spike-evoked adaptation current gives a more
realistic description to IF models (including a physiological
basis of the widespread phenomenon of \textit{spike-frequency
adaptation}), without a substantial increase of complexity. In
detail, we consider that, during subthreshold integration, the
membrane potential evolves according to

\begin{equation}\label{eq1}
   \frac{dV}{dt} = - \frac{(V-V_{\rm rest})}{\tau_{\rm m}} + I_{\rm adapt} + \mu
   + \xi(t),
\end{equation}

\noindent where the time constant given by the leaky current is
$\tau_{\rm m}$, the resting potential is $V_{\rm rest}$, the
adaptation current is denoted by $I_{\rm adapt}$, and the external
signal is represented by the constant drift $\mu$. Randomness
arises at different stages during signal transduction and neuronal
communication \cite{Burkitt2006}, and here it is included simply
as an additive Gaussian white noise, $\langle \xi(t) \rangle = 0$
and $\langle \xi(t) \xi(t') \rangle = 2D\delta(t'-t)$. This
dynamics governs the evolution of the membrane potential during
the subthreshold integration; whenever the potential reaches a
certain threshold $V_{\rm thr}$, a spike is produced (or declared)
and immediately after, the potential is set to a reset value
$V_{\rm r}$. By a simple re-scaling of $V$, we can consider
$V_{\rm rest} = 0$ and $V_{\rm thr} = 1$ without loss of
generality. Furthermore, for simplicity we define $V_{\rm r} =
V_{\rm rest} = 0$.

One of the most important output properties of a spiking neuron is
its firing rate, i.e., the number of spikes produced in certain
time. When analyzed as a function of an input feature, the
resulting \textit{tuning} function represents a input-output
mapping, which usually is modulated by different factors. For the
deterministic LIF model, spikes can be produced only when the
constant drift is above certain value, see fig.~\ref{fig.1}(a),
which separates two different regimes: sub- and supra-threshold
regimes. This discontinuous mapping is monotonically smoothed by
noise, see figs.~\ref{fig.1}(a) and \ref{fig.1}(b). In the
subthreshold mode, the input drift drives the neuronal dynamics
towards a quiescent state below the potential threshold, see
fig.~\ref{fig.1}(c) (upper trace in light-gray line), and noise is
essential to produce any spike. In the supra-threshold mode, the
dynamical state is set to a repetitive firing regime, see
fig.~\ref{fig.1}(c) (lower trace in dark-gray line), and noise has
no such a fundamental role. Interestingly, a given firing rate can
be obtained in both regimes by proper combinations of input
parameters, $(\mu,D)$, see fig.\ref{fig.1}(b).

A spike-evoked adaptation current mimics the effects of
Ca$^{2+}-$activated K$^{+}$ or \textit{after-hyperpolarization}
currents \cite{MadisonNicoll1984}, which are widely expressed in
the mammalian nervous system \cite{Sah1996}, and can be minimally
modelled by a current-based description \cite{LiuWang2001,
Benda_etal_2010, Urdapilleta2011, Muller_etal_2007}, $I_{\rm
adapt}(t) = -g_{\rm adapt} ~ x(t)$, where the adaptation process
$x(t)$ filters the output spike train according to

\begin{equation}\label{eq2}
    \frac{dx}{dt} = - \frac{x}{\tau_{\rm a}} + \alpha \sum_i
    \delta(t-t_i).
\end{equation}

In this equation, $\tau_{\rm a}$ and $\alpha$ define the temporal
and the output scales of the adaptation process, respectively, and
$\delta(t_i)$ is a pulse (Dirac delta function) representing a
spike occurring at time $t_i$. Without loss of generality
\cite{Urdapilleta2011}, $g_{\rm adapt}$ can be conveniently
re-scaled so the temporal profile of the adaptation current during
the input integration of the $n$-th interspike interval,
$\tau_{\rm n} = t_{\rm n+1} - t_{\rm n}$, reads

\begin{equation}\label{eq3}
   I_{\rm adapt} = - \frac{\varepsilon_{\rm n}}{\tau_{\rm a}}
   \exp[-(t-t_{\rm n})/\tau_{\rm a}].
\end{equation}

Coupling between successive initial adaptation strengths is
provided by integration of eq.~(\ref{eq2}) during the arrival of a
new spike,

\begin{equation}\label{eq4}
   \varepsilon_{\rm n+1} = \varepsilon_{\rm n}
   \exp(-\tau_{\rm n}/\tau_{\rm a}) + \alpha,
\end{equation}

\noindent see fig.~\ref{fig.1}(c). As depicted schematically in
fig.~\ref{fig.1}(d), the preceding relationship supports a
history-dependent process that creates correlations between
subsequent ISIs \cite{Urdapilleta2011}. The statistics of the
($n+1$)-th ISI, $\tau_{{\rm n}+1}$, is conditioned to the level of
the ($n+1$)-th initial adaptation strength, $\varepsilon_{{\rm
n}+1}$, through a temporally inhomogeneous first-passage-time
problem with an exponential time-dependent drift
\cite{Urdapilleta2011b, Urdapilleta2012, Urdapilleta2015}.
Furthermore, since the initial adaptation strength of the
($n+1$)-th period depends exclusively on both variables at the
immediately previous stage, $\varepsilon_{{\rm n}}$ and
$\tau_{{\rm n}}$, through eq.~(\ref{eq4}), this scheme constitutes
a Markov process with both an observable (interspike interval) and
a hidden variable (initial adaptation strength). However,
correlations between ISIs are not limited to consecutive periods,
but extends to all previous outcomes, due to a nested dependence
via the hidden variable. A similar HMM can be defined for other
history-dependent processes that also produce spike-frequency
adaptation and generate correlations between ISIs
\cite{Chacron_etal_2003, Nesse_etal_2010, SchwalgerLindner2010}.

\section{Correlations in the sequence of interspike intervals}

To quantify these correlations it is useful to define the serial
correlation coefficient (SCC), which, in stationary conditions, is
given by

\begin{equation}\label{eq5}
   \rho_{\rm k} = \frac{\langle \tau_{\rm n}\tau_{\rm n+k} \rangle -
   \langle \tau \rangle^2}{\langle (\tau - \langle \tau \rangle)^2 \rangle},
\end{equation}

\noindent where brackets denote ensemble average and k is the lag
between successive ISIs. To compute the SCCs at any lag, it is
necessary to quantitatively describe the HMM depicted in
fig.~\ref{fig.1}(d). Given the deterministic update defined by
eq.~(\ref{eq4}), this HMM is completely characterized by the
transition probability density

\begin{eqnarray}\label{eq6}
    f(\varepsilon_{\rm n},\tau_{\rm n}|\varepsilon_{\rm n-1},\tau_{\rm
    n-1}) = \delta\{\varepsilon_{\rm n} -
    [\varepsilon_{\rm n-1}\exp(-\tau_{\rm n-1}/\tau_{\rm
    a})+\alpha]\}\nonumber\\
    \times~\phi(\tau_{\rm n}|\varepsilon_{\rm n}),\hspace{3.5cm}
\end{eqnarray}

\noindent where $\phi(\tau_{\rm n}|\varepsilon_{\rm n})$ is the
ISI probability density for the (temporally inhomogeneous) system
defined by eqs.~(\ref{eq1}) and (\ref{eq3}) evolving from the
reset to the threshold, conditioned to the explicit knowledge of
the initial strength $\varepsilon_{\rm n}$. Recently, we showed
that this probability density can be expressed as a series
expansion in terms of the initial strength of the adaptation
current for any one-dimensional IF model,

\begin{equation}\label{eq7}
    \phi(\tau|\varepsilon) = \sum_{\rm n=0}^{\infty} \varepsilon^{\rm
    n}~\phi_{\rm n}(\tau),
\end{equation}

\noindent and, particularly, we explicitly computed all terms for
the LIF model \cite{Urdapilleta2015}.

\begin{figure}[!t]
\centerline{\includegraphics[width=0.95\linewidth]{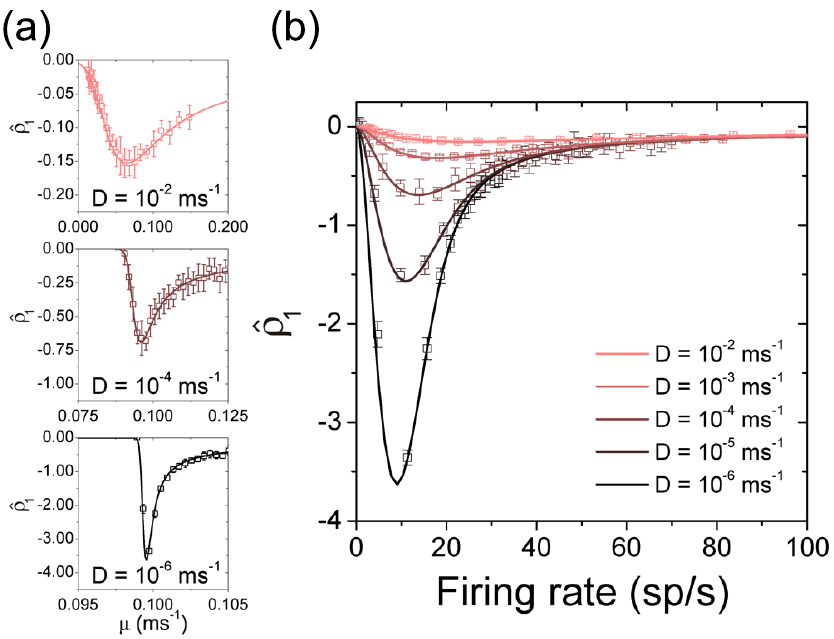}}
\caption{(Color online) Normalized SCC at lag $1$, $\hat{\rho}_1 =
\rho_1 / \alpha$, when spiking rate responses are set by the drift
$\mu$. (a) $\hat{\rho}_1$ as a function of the constant input
$\mu$ driving the neuron, for different noise intensities.
Analytical expressions (colored lines) and correlations obtained
from the simulated dynamics (corresponding symbols, each
calculated from a sequence of $N = 10^{6}$ ISIs) show an excellent
agreement. (b) $\hat{\rho}_1$ represented as a function of the
firing rate, for different noise intensity scenarios.
Interestingly, the largest absolute value is reached at
approximately the same firing rate in all cases. Parameters: Same
as in fig. 1, except that $\alpha = 0.01$ in simulations and
$\tau_{\rm a} = 100~{\rm ms}$.} \label{fig.2}
\end{figure}

Based on the transition probability defined by eq.~(\ref{eq6}), in
a previous work \cite{Urdapilleta2011}, we computed $\langle
\tau_{\rm n}\tau_{\rm n+k} \rangle$ and $\rho_{\rm k}$, assuming
that the statistics given by eq.~(\ref{eq7}) and all nested
expressions produced when computing SCCs at higher lags are
developed up to order $1$. In this small adaptation scenario (set
by small values of $\alpha$), correlations for any IF model
develop a geometrical structure,

\begin{equation}\label{eq8}
   \rho_{\rm k} = \left[ \tilde{\phi}^L_0 (1/\tau_{\rm a}) \right]^{\rm k - 1}
   ~\rho_1,
\end{equation}

\noindent where $\tilde{\phi}^L_0 (s)$ is the Laplace transform of
the ISI density function for the unperturbed system (i.e., without
the adaptation current), and $\rho_1$, the first SCC, reads

\begin{eqnarray}\label{eq9}
   \rho_1 = -\alpha ~ \frac{\langle \tau \rangle_{\phi_1}}
   {\left[ 1-\tilde{\phi}^L_0(1/\tau_{\rm a})\right]\langle
   (\tau - \langle \tau \rangle)^2\rangle_{\phi_0}}\nonumber\\
   \times\left[ \tilde{\phi}^L_0(1/\tau_{\rm a})~\langle \tau
   \rangle_{\phi_0} + \frac{d \tilde{\phi}^L_0(s)}{ds}
   \rfloor_{1/\tau_{\rm{a}}}\right].
\end{eqnarray}

The indexed brackets in eq.~(\ref{eq9}) are the contributions to
the moments computed with the functions indicated by the
respective index,

\begin{equation}\label{eq10}
    \langle \tau \rangle_{\phi_{\rm n}} = \int_0^{\infty}
    \tau~\phi_{\rm n}(\tau)~d\tau = - \frac{d\tilde{\phi}^L_{\rm
    n}(s)}{ds}\rfloor_{s=0}.
\end{equation}

Therefore, within this framework, the two quantities needed to
evaluate all SCCs are the unperturbed ISI density function
(expressed in the Laplace domain) $\tilde{\phi}^L_0(s)$ and the
first order correction $\tilde{\phi}^L_1(s)$ (or, at least, its
effect on the mean ISI, $\langle \tau \rangle_{\phi_1}$). For the
LIF model, these quantities read

\begin{equation}\label{eq11}
    \tilde{\phi}^L_0(s) = {\rm e}^{-\left( Z_{\rm thr}^2 -
    Z_{\rm r}^2 \right)/4}~\frac{\mathcal{D}_{-\tau_{\rm m} s}(Z_{\rm r})}
    {\mathcal{D}_{-\tau_{\rm m} s}(Z_{\rm thr})},
\end{equation}

\begin{eqnarray}\label{eq12}
    \tilde{\phi}^L_1(s) = \frac{\sqrt{\tau_{\rm m}/D}}
    {1-\tau_{\rm d}/\tau_{\rm m}}~\frac{{\rm e}^{-\left( Z_{\rm thr}^2
    - Z_{\rm r}^2 \right)/4}}{\mathcal{D}_{-\tau_{\rm m} s}
    (Z_{\rm thr})}~s~\Big[ \mathcal{D}_{-\tau_{\rm m}(s+1/\tau_{\rm m})}
    (Z_{\rm r}) \nonumber\\
    - \frac{\mathcal{D}_{-\tau_{\rm m}(s+1/\tau_{\rm m})}(Z_{\rm thr})}
    {\mathcal{D}_{-\tau_{\rm m}(s+1/\tau_{\rm d})}(Z_{\rm thr})}~
    \mathcal{D}_{-\tau_{\rm m}(s+1/\tau_{\rm d})}(Z_{\rm r})
    \Big],\hspace{0.5cm}
\end{eqnarray}

\noindent where $Z_{\rm i} = \sqrt{\tau_{\rm m}/D}~(\mu-V_{\rm
i}/\tau_{\rm m})$, and $\mathcal{D}_{\nu}(z)$ is the parabolic
cylinder function according to Whittaker's notation
\cite{Handbook2010}.

\begin{figure}[!t]
\centerline{\includegraphics[width=0.95\linewidth]{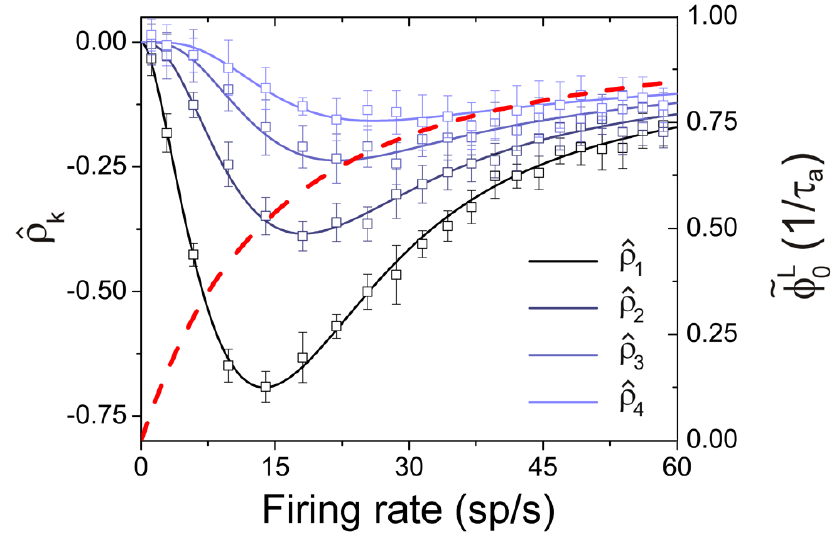}}
\caption{(Color online) Normalized SCCs at higher lags,
$\hat{\rho}_{\rm k} = \rho_{\rm k} / \alpha$. Analytical
expressions (bluish colored lines) perfectly agree with data
obtained from simulations (corresponding symbols, each calculated
from a sequence of $N = 5\times 10^{6}$ ISIs). The factor defining
the geometrical structure, $\tilde{\phi}_0^L(1/\tau_{\rm a})$, is
also shown (red dashed line, scale on the right side). As in fig.
2, firing rate was determined by varying the constant drift $\mu$.
Parameters as in fig. 2, with $D = 10^{-4}~{\rm ms}^{-1}$.}
\label{fig.3}
\end{figure}

In fig.~\ref{fig.2}(a) we show the first SCC (normalized by
$\alpha$), which sets the basis for all other SCCs at higher lags,
as a function of the constant input driving the spiking dynamics,
$\mu$, for different noise intensities. When $\rho_{1}$ is
represented as a function of the firing rate elicited by the
constant input, fig.~\ref{fig.2}(b), we can observe that the
behavior is approximately conserved, but scaled, across the
different cases. Importantly, irrespective of the noise intensity,
$\rho_1$ exhibits a minimum around certain firing frequency.

The geometrical structure of the SCCs at higher lags,
eq.~(\ref{eq8}), depends on a scaling factor given by the ISI
density function of the system without adaptation, $\phi_0(\tau)$,
but Laplace-transformed and evaluated at a specific value,
$\tilde{\phi}^L_0 (1/\tau_{\rm a})$. In fig.~\ref{fig.3} we show
the first $4$ correlation coefficients, for a representative case,
as a function of the firing rate elicited by varying the constant
input current, $\mu$. The scaling factor is shown in red dashed
line, whose scale can be read on the right margin. Given the
monotonic character of this scaling factor, successive minima
slightly shift towards higher firing rates as lag increases.

\begin{figure*}[!t]
\centerline{\includegraphics[width=0.7125\textwidth]{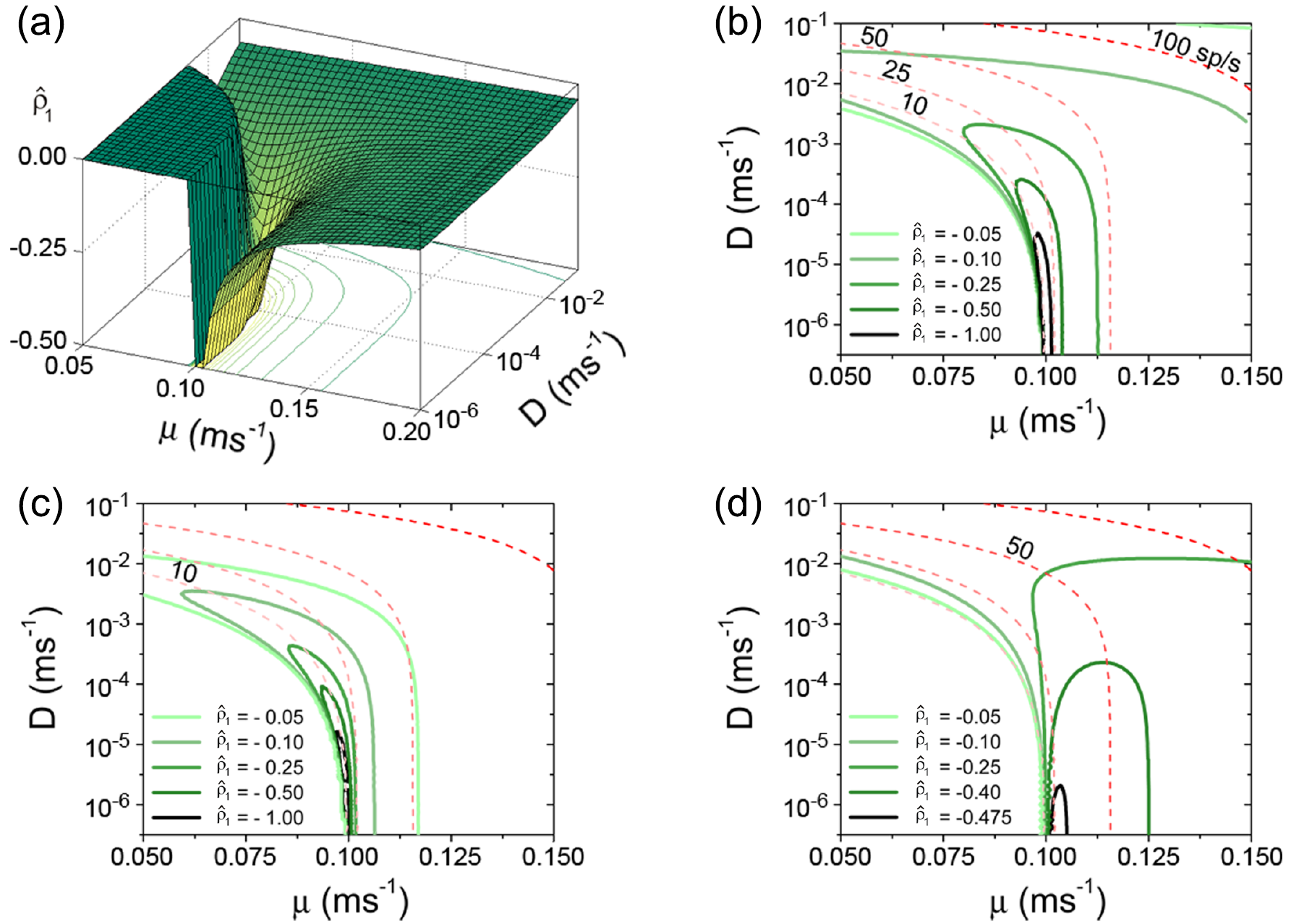}}
\caption{(Color online) Structure of correlations. (a) Normalized
SCC at lag $1$, $\hat{\rho}_1$ in parameter space. A deep valley
approximately follows the trajectory of the isoline corresponding
to a specific firing rate (compare normalized SCC isolines at the
bottom plane with firing rate isolines in fig. 1(b)). (b)
Bi-dimensional representation of the contour structure. Greenish
lines represent contours of $\hat{\rho}_1$ (indicated in the
inset), whereas reddish dashed lines are isolines of firing rates
(indicated \textit{in situ}). Correlation contours approximately
follow the isoline corresponding to $10~{\rm sp/s}$ (i.e., $\sim
1/\tau_{\rm a}$). Parameters for (a) and (b): same as in fig. 2,
including $\tau_{\rm a} = 100~{\rm ms}$. (c) The structure of
correlations develops narrowly around a lower firing rate ($\sim
2~{\rm sp/s}$) when time scale of the adaptation current is
increased ($\tau_{\rm a} = 500~{\rm ms}$). (d) The opposite is
true when time scale is reduced. Overall, correlation weakens (see
inset) and develops loosely around the indicated firing rate. In
this case, $\tau_{\rm a} = 20~{\rm ms}$ and $\sim 50~{\rm sp/s}$.}
\label{fig.4}
\end{figure*}

\section{Noise-induced correlations}
Overall, the preceding results are very similar to those we have
previously obtained for a \textit{perfect} IF (PIF) neuron model
\cite{Urdapilleta2011}. In this study, we have shown the same
behavior, but analyzed as a function of the constant input $\mu$,
which actually is equivalent to the firing rate but scaled (in the
pure PIF model, noise does not modulate the firing rate). Even
when it is useful to gain theoretical insight with a tractable
model, the PIF model lacks of biological realism, as it only can
be set in the supra-threshold regime and noise simply randomizes
spike times without any fundamental role. Here, with the study of
the LIF model, we can focus on the sub-threshold regime and
analyze the contribution of noise in creating correlations.

Given that, according to fig.~\ref{fig.2}(b), minima of
correlations for different cases are set around certain firing
rate, the key idea to explore is whether there is a matching of
time scales between the adaptation process, which is the
responsible for creating correlations, and the firing state. In
fig.~\ref{fig.4}(a) we show the first SCC in the input parameter
space, $\mu$ and $D$. In the low-noise limit, it can be observed a
deep valley, characterized by the value of $\mu$ that elicits a
particular firing rate. As noise increases, the position of this
valley moves along, bending towards the $D$-axis. If we focus on
the contour levels, we can distinguish that $\rho_1$-isolines are
similar to those of the firing rate, see fig.~\ref{fig.1}(b),
implying that the development of strong correlations are
concomitant to a particular firing rate. This is further developed
in fig.~\ref{fig.4}(b), where the contour levels of $\rho_1$ and
the firing rate are plotted together in the parameter space. As a
general trend, correlations are structured around $10~{\rm sp/s}$,
which tentatively corresponds to $1/\tau_{\rm a}$. Since the
adaptation process is the responsible for creating a
history-dependent spike train, its time constant sets the scale in
which spikes should be produced to maximize the influence of the
update rule, eq.~(\ref{eq4}), on the development of correlations.
Therefore, by changing the adaptation time constant we should
observe that correlations develop around a different contour level
of the firing rate. This scenario is shown in figs.~\ref{fig.4}(c)
and \ref{fig.4}(d), where adaptation time scale has been set at
$\tau_{\rm a} = 500~{\rm ms}$ and $\tau_{\rm a} = 20~{\rm ms}$,
respectively. As expected, correlations organize around $2~{\rm
sp/s}$ (shown as below $10~{\rm sp/s}$) and $50~{\rm sp/s}$,
respectively, with the additional effect that they are
strengthened (weakened) as adaptation time scale increases
(decreases).

\begin{figure}[!t]
\centerline{\includegraphics[width=0.95\linewidth]{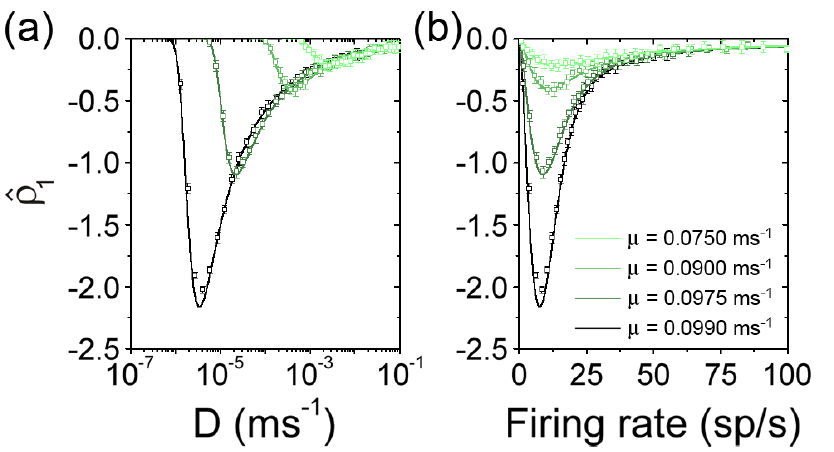}}
\caption{(Color online) Noise-induced correlations. (a) Normalized
SCC at lag $1$, $\hat{\rho}_1$, as a function of the noise
intensity, for different sub-threshold scenarios (established by
$\mu$). The transition to the supra-threshold condition occurs at
$\mu = 0.100~{\rm ms}^{-1}$ (parameters as in fig. 2). Greenish
lines represent analytical expressions, whereas corresponding
symbols are data obtained from simulations (each data-point was
calculated from a sequence of $N = 5 \times 10^{6}$ ISIs). Extreme
correlations occur at a finite noise, whose value depends on the
specific sub-threshold case. (b) When represented as a function of
the ongoing firing rate, all minima are set around the same rate,
indicative of a matching of time scales sustained by noise.}
\label{fig.5}
\end{figure}

The preceding observation corresponds to a matching of time
scales: for a given adaptation process (a defined $\tau_{\rm a}$),
correlations are stronger when the neuron fires at a certain
firing rate. In particular, this firing rate can be elicited in a
subthreshold regime and, furthermore, be driven by noise. In this
case, for example, the constant input $\mu$ may be fixed by
external influences and noise can be considered as a parameter.
Different cases, corresponding to different values of $\mu$, are
shown in fig.~\ref{fig.5}(a). The closer the value of $\mu$ to the
critical value separating sub- and supra-threshold regimes (here,
$0.10~{\rm ms}^{-1}$), the stronger the correlations and, of
course, the weaker the noise intensity that maximizes them.
However, as we argue above, the intrinsic phenomenon is a matching
of time scales, so when represented as a function of the firing
rate elicited by the noise, see fig.~\ref{fig.5}(b), all cases
display their maximum of correlations at the same firing rate.
From a different perspective, for a given system (i.e., a defined
$\tau_{\rm a}$ and $D$), there will be certain subthreshold
external input $\mu$ that produces the strongest negative
correlations (sustained by noise).

A minimum in the first SCC as a function of the noise intensity
was previously reported for a related model
\cite{Chacron_etal_2003}. In this study, the authors have
numerically found a shallow minimum in $\rho_1$ at a finite noise,
for a LIF neuron model with a history-dependent threshold. This
minimum was not very pronounced probably because the system was
set in the supra-threshold regime. At the light of our results, a
precise value of the noise intensity will be influential only in a
sub-threshold condition.

\section{Influence on spike-count statistics}
The development of correlations between ISIs has an important
impact on rate coding. In general, the \textit{Fano factor} is
utilized to characterize the relative importance of the first two
moments of the statistics defined by the number of spikes observed
in a temporal window of length $T$, as ${\rm FF}_T = \langle
\Delta n_T^2 \rangle / \langle n_T \rangle$, where $\langle n_T
\rangle$ and $\langle \Delta n_T^2 \rangle$ are the mean and the
variance, respectively. For $T\rightarrow \infty$, the Fano factor
converges to \cite{CoxLewis1966}

\begin{equation}\label{eq12}
    {\rm FF}_{\infty} = {\rm lim}_{T\rightarrow \infty}
    ~ {\rm FF}_T = {\rm CV}^2~\left( 1 +
    2~\sum_{{\rm k}=1}^{\infty} \rho_{\rm k} \right),
\end{equation}

\noindent where ${\rm CV}$ is the \textit{coefficient of
variation}, defined on the statistics of single ISIs as ${\rm CV}
= \sqrt{\langle \Delta \tau^2 \rangle} / \langle \tau \rangle$.

\begin{figure}[!t]
\centerline{\includegraphics[width=0.95\linewidth]{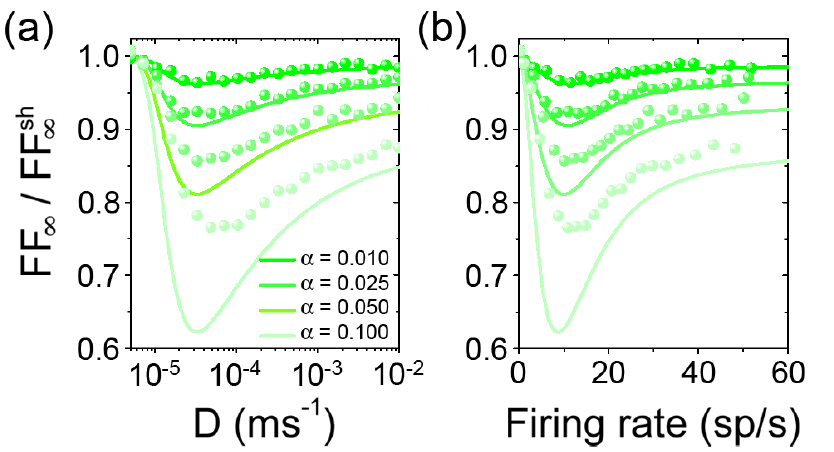}}
\caption{(Color online) Spike train regularization by negative
correlations. (a) Ratio between the Fano factor and the Fano
factor of a surrogate spike train with ISI correlations removed by
shuffling, ${\rm FF}_{\infty} / {\rm FF}_{\infty}^{\rm sh}$, as a
function of the noise intensity, for a LIF neuron model set in a
subthreshold regime (symbols). This ratio highlights the exclusive
contribution of correlations on the reduction of the spike-train
variability, $1 + 2~\sum_{{\rm k}=1}^{\infty} \rho_{\rm k}$
(lines). Temporal windows used to compute spike-counts were large
enough to assure asymptotic conditions. Parameters as in fig. 2,
with $\mu = 0.0975~{\rm ms}^{-1}$. Different contributions of the
adaptation current to the dynamics are considered, see different
values of $\alpha$. (b) When represented as a function of the
evoked firing rate, the strongest reduction of the spike-count
variability is observed at a particular firing rate.}
\label{fig.6}
\end{figure}

Therefore, a process that creates negative correlations also
generates a reduction of the spike-count variability, as the
factor $1 + 2~\sum_{{\rm k}=1}^{\infty} \rho_{\rm k}$ is less than
unity. These correlations can be removed from a spike train by
\textit{shuffling} the order of the ISIs, a procedure that creates
a surrogate spike train with exactly the same single ISI
statistics, but no correlations between them. Then, the ratio
${\rm FF}_{\infty} / {\rm FF}_{\infty}^{\rm sh}$ clearly
highlights the exclusive contribution of correlations in the
reduction of the spike-count variability. In fig.~\ref{fig.6}(a)
we show this reduction as a function of the noise intensity, in a
given subthreshold regime. When represented as a function of the
firing rate evoked, fig.~\ref{fig.6}(b), the strongest reduction
of the spike-count variability is positioned at the firing rate
that matches the adaptation process ($10$~Hz).

In fig.~\ref{fig.6}, we can observe that our analytical derivation
perfectly agrees to numerical results in the perturbative regime
(brighter green symbols), where this framework is valid, whereas
higher order effects on correlations are visible when $\alpha$
increases (dimer green symbols). Clearly, these effects have to
oppose the linear decrease in order to prevent an unlimited growth
beyond physical significance. However, the conclusion that a
reduction of the spike-count variability and, therefore, a
regularization of the spike train is maximized at a specific
firing frequency (here, sustained by a specific value of noise
intensity) holds well beyond the perturbative scenario, indicating
that the effect subsists in more realistic adaptation conditions.

\section{Conclusions}
In this work we have analyzed the development of negative
correlations in a LIF neuron model with a spike-triggered
adaptation current. This system is adequate to explore both supra-
and sub-threshold regimes. Whereas the first regime has been
previously studied, and also confirmed by the present approach,
the second one has been never characterized. We found that
correlations are stronger as the neuron fires at a particular
firing rate, defined by the inverse of the adaptation time scale.
Obviously, this scenario can be obtained in both firing regimes,
and when restricted to the sub-threshold case, noise plays a
fundamental role, by driving the specific firing rate that
maximizes correlations. Given that the sum of negative
correlations at different lags produces a regularization of the
long-term spike-count variability \cite{Chacron_etal_2001,
Urdapilleta2011, CoxLewis1966}, this noise-sustained effect
implies that noise may have a constructive role in neural rate
codes when adaptation currents are present.

\acknowledgments This work was supported by the Consejo de
Investigaciones Cient\'ificas y T\'ecnicas de la Rep\'ublica
Argentina.


\begin{thebibliography}{0}

\bibitem{Gerstner_etal_2014}
  \Name{Gerstner W., Kistler W. M., Naud R. \and Paninski L.}
  \Book{Neuronal dynamics: From single neurons to networks and models of cognition}
  \Publ{Cambridge University Press, Cambridge}
  \Year{2014}.

\bibitem{Perkel_etal_1967}
  \Name{Perkel D. H., Gerstein G. L. \and Moore G. P.}
  \REVIEW{Biophys. J.}{7}{1967}{391}.

\bibitem{ShadlenNewsome1998}
  \Name{Shadlen M. N. \and Newsome W. T.}
  \REVIEW{J. Neurosci.}{18}{1998}{3870}.

\bibitem{LowenTeich1992}
  \Name{Lowen S. B. \and Teich M. C.}
  \REVIEW{J. Acoust. Soc. Am.}{92}{1992}{803}.

\bibitem{RatnamNelson2000}
  \Name{Ratnam R. \and Nelson M. E.}
  \REVIEW{J. Neurosci.}{20}{2000}{6672}.

\bibitem{NeimanRussell2001}
  \Name{Neiman A. \and Russell D. F.}
  \REVIEW{Phys. Rev. Lett.}{86}{2001}{3443}.

\bibitem{Bahar_etal_2001}
  \Name{Bahar S., Kantelhardt J. W., Neiman A., Rego H. H. A., Russell D. F., Wilkens L., Bunde A. \and Moss F.}
  \REVIEW{Europhys. Lett.}{56}{2001}{454}.

\bibitem{Nawrot_etal_2007}
  \Name{Nawrot M. P., Boucsein C., Rodriguez-Molina V., Aertsen A., Gr\"un S. \and Rotter S.}
  \REVIEW{Neurocomputing}{70}{2007}{1717}.

\bibitem{Farkhooi_etal_2009}
  \Name{Farkhooi F., Strube-Bloss M. F. \and Nawrot M. P.}
  \REVIEW{Phys. Rev. E}{79}{2009}{021905}.

\bibitem{Peterson_etal_2014}
  \Name{Peterson A. J., Irvine D. R. F. \and Heil P.}
  \REVIEW{J. Neurosci.}{34}{2014}{15097}.

\bibitem{AkerbergChacron2011}
  \Name{Avila-Akerberg O. \and Chacron M. J.}
  \REVIEW{Exp. Brain Res.}{210}{2011}{353}.

\bibitem{LiuWang2001}
  \Name{Liu Y. H. \and Wang X. J.}
  \REVIEW{J. Comput. Neurosci.}{10}{2001}{25}.

\bibitem{Chacron_etal_2001}
  \Name{Chacron M. J., Longtin A. \and Maler L.}
  \REVIEW{J. Neurosci.}{21}{2001}{5328}.

\bibitem{PrescottSejnowski2008}
  \Name{Prescott S. A. \and Sejnowski T. J.}
  \REVIEW{J. Neurosci.}{28}{2008}{13649}.

\bibitem{Benda_etal_2010}
  \Name{Benda J., Maler L. \and Longtin A.}
  \REVIEW{J. Neurophysiol.}{104}{2010}{2806}.

\bibitem{Schwalger_etal_2010}
  \Name{Schwalger T., Fisch K., Benda J. \and Lindner B.}
  \REVIEW{PLoS Comput. Biol.}{6(12)}{2010}{e1001026}.

\bibitem{Nesse_etal_2010}
  \Name{Nesse W. H., Maler L. \and Longtin A.}
  \REVIEW{Proc. Natl. Acad. Sci. USA}{107(51)}{2010}{21973}.

\bibitem{Farkhooi_etal_2011}
  \Name{Farkhooi F., Muller E. \and Nawrot M. P.}
  \REVIEW{Phys. Rev. E}{83}{2011}{050905}.

\bibitem{Urdapilleta2011}
  \Name{Urdapilleta E.}
  \REVIEW{Phys. Rev. E}{84}{2011}{041904}.

\bibitem{SchwalgerLindner2013}
  \Name{Schwalger T. \and Lindner B.}
  \REVIEW{Front. Comput. Neurosci.}{7}{2013}{164}.

\bibitem{Shiau_etal_2015}
  \Name{Shiau L., Schwalger T. \and Lindner B.}
  \REVIEW{J. Comput. Neurosci.}{38}{2015}{589}.

\bibitem{CoxLewis1966}
  \Name{Cox D. R. \and Lewis P. A. W.}
  \Book{The statistical analysis of series of events}
  \Publ{Methuen \& Co., Ltd., London}
  \Year{1966}.

\bibitem{VanVreeswijk2010}
  \Name{van Vreeswijk C.}
  \Book{{\rm in} Analysis of parallel spike trains}
  \Editor{Gr\"un S. \and Rotter S.}
  \Publ{Springer-Verlag, Berlin}
  \Year{2010}
  \Page{3}.

\bibitem{Chacron_etal_2004}
  \Name{Chacron M. J., Lindner B. \and Longtin A.}
  \REVIEW{Phys. Rev. Lett.}{92}{2004}{080601}.

\bibitem{Lindner_etal_2005}
  \Name{Lindner B., Chacron M. J. \and Longtin A.}
  \REVIEW{Phys. Rev. E}{72}{2005}{021911}.

\bibitem{Prescott_etal_2008}
  \Name{Prescott S. A., De Koninck Y. \and Sejnowski T. J.}
  \REVIEW{PLoS Comput. Biol.}{4}{2008}{e1000198}.

\bibitem{MatoSamengo2008}
  \Name{Mato G. \and Samengo I.}
  \REVIEW{Neural Comput.}{20}{2008}{2418}.

\bibitem{Rauch_etal_2003}
  \Name{Rauch A., La Camera G., L\"uscher H. -R., Senn W. \and Fusi S.}
  \REVIEW{J. Neurophysiol.}{90}{2003}{1598}.

\bibitem{LaCamera_etal_2008}
  \Name{La Camera G., Giugliano M., Senn W. \and Fusi S.}
  \REVIEW{Biol. Cybern.}{99}{2008}{279}.

\bibitem{VanVreeswijkSompolinsky1998}
  \Name{van Vreeswijk C. \and Sompolinsky H.}
  \REVIEW{Neural Comp.}{10}{1998}{1321}.

\bibitem{VogelsAbbott2009}
  \Name{Vogels T. P. \and Abbott L. F.}
  \REVIEW{Nat. Neurosci.}{12}{2009}{483}.

\bibitem{Chacron_etal_2003}
  \Name{Chacron M. J., Pakdaman K. \and Longtin A.}
  \REVIEW{Neural Comp.}{15}{2003}{253}.

\bibitem{HutcheonYarom2000}
  \Name{Hutcheon B. \and Yarom Y.}
  \REVIEW{Trends Neurosci.}{23(5)}{2000}{216}.

\bibitem{BrownAdams1980}
  \Name{Brown D. A. \and Adams P. R.}
  \REVIEW{Nature}{283}{1980}{673}.

\bibitem{Nesse_etal_2008}
  \Name{Nesse W. H., Del Negro C. A. \and Bressloff P. C.}
  \REVIEW{Phys. Rev. Lett.}{101}{2008}{088101}.

\bibitem{Pozzorini_etal_2013}
  \Name{Pozzorini C., Naud R., Mensi S. \and Gerstner W.}
  \REVIEW{Nat. Neurosci.}{16}{2013}{942}.

\bibitem{Burkitt2006}
  \Name{Burkitt A. N.}
  \REVIEW{Biol. Cybern.}{95}{2006}{1}.

\bibitem{MadisonNicoll1984}
  \Name{Madison D. V. \and Nicoll R. A.}
  \REVIEW{J. Physiol.}{354}{1984}{319}.

\bibitem{Sah1996}
  \Name{Sah P.}
  \REVIEW{Trends Neurosci.}{19}{1996}{150}.

\bibitem{Muller_etal_2007}
  \Name{Muller E., Buesing L., Schemmel J. \and Meier K.}
  \REVIEW{Neural Comput.}{19}{2007}{2958}.

\bibitem{Urdapilleta2011b}
  \Name{Urdapilleta E.}
  \REVIEW{Phys. Rev. E}{83}{2011}{021102}.

\bibitem{Urdapilleta2012}
  \Name{Urdapilleta E.}
  \REVIEW{J. Phys. A: Math. Theor.}{45}{2012}{185001}.

\bibitem{Urdapilleta2015}
  \Name{Urdapilleta E.}
  \REVIEW{J. Phys. A: Math. Theor.}{48}{2015}{505001}.

\bibitem{SchwalgerLindner2010}
  \Name{Schwalger T. \and Lindner B.}
  \REVIEW{Eur. Phys. J. Special Topics}{187}{2010}{211}.

\bibitem{Handbook2010}
  \Name{Olver F. W. J., Lozier D. W., Boisvert R. F. \and Clark C. W. (Editors)}
  \Book{NIST Handbook of Mathematical Functions}
  \Publ{Cambridge University Press, New York}
  \Year{2010}.

\end{thebibliography}
\end{document}